\documentclass[rnote]{aa} 

\usepackage{graphicx}
\usepackage{txfonts}
\usepackage{color}
\newcommand{\kms}{km\,s$^{-1}$}       

\newcommand{\vlsr}{$\upsilon_{\rm LSR}$}        

\newcommand{\tadv}{$\int \! T_{\rm A} d\upsilon$}

\newcommand{\trms}{$T_{\rm rms}$}

\newcommand{\molh}{H$_{2}$}                              

\newcommand{\water}{H$_{2}$O}

\newcommand{\peroxy}{H$_{2}$O$_{2}$}  
\newcommand{\peroxyo}{HOOH}

\newcommand{\molo}{O$_{2}$}                     

\newcommand{\about}{$\sim$}                       
\newcommand{\powten}[1]{10$^{#1}$}
\newcommand{\ammonia}{{\rm NH}$_3$}     

\newcommand{\roa}{$\rho \, {\rm Oph \, A}$}

\newcommand{\asec}{$^{\prime \prime}$}

\newcommand{\atwozero}{$\alpha_{2000}$}
\newcommand{\dtwozero}{$\delta_{2000}$}

\newcommand{\radot}[4]{\mbox{#1$^{\rm h}$#2$^{\rm m}$#3$\stackrel {\rm s}{_{\bf\cdot}}$#4}}  

\newcommand{\decdot}[4]{\mbox{#1$^{\circ}$ #2$^{\prime}$ #3$\stackrel {\prime \prime}{_{\bf \cdot}}$#4}}

\begin{document}

\title{Search for HOOH in Orion
\thanks{Based on observations with APEX, which is a 12\,m diameter submillimetre telescope at 5100\,m altitude on Llano Chajnantor in Chile. The telescope is operated by Onsala Space Observatory, Max-Planck-Institut f\"ur Radioastronomie (MPIfR), and European Southern Observatory (ESO).}
}


\author{
                R. Liseau\inst{1}                                              
        \and
                B. Larsson\inst{2}
         }

  \institute{Department of Earth and Space Sciences, Chalmers University of Technology, Onsala Space Observatory, SE-439 92 Onsala, Sweden, 
              \email{\small{rene.liseau@chalmers.se}} 
        \and 
                AlbaNova University Centre, Stockholm University, Department of Astronomy, SE-106 91 Stockholm, Sweden 
        }

\date{Received ... / Accepted ...}
%
%

\abstract
{The abundance of key molecules determines the level of  cooling that is necessary for the formation of stars and planetary systems. In this context, one needs to understand the details of the time dependent oxygen chemistry, leading to the formation of  \molo\ and \water.}
{We aim to determine the degree of correlation between the occurrence of \molo\ and \peroxyo\ (hydrogen peroxide)  in star-forming molecular clouds. We  first detected \molo\ and \peroxyo\ in \roa, we now search for \peroxyo\ in Orion OMC\,A, where \molo\ has also been detected.}
{We mapped a $3^{\prime} \times 3^{\prime}$ region around Orion \molh-Peak\,1 with the Atacama Pathfinder Experiment (APEX). In addition to several maps in two transitions of  \peroxyo, viz. 219.17\,GHz and 251.91\,GHz, we obtained single-point spectra for another three transitions towards the position of maximum emission.}
{Line emission at the appropriate LSR-velocity (Local Standard of Rest) and at the level of $\ge 4 \sigma$ was found for two transitions, with lower S/N ($2.8 - 3.5 \sigma$) for another two transitions, whereas for the remaining transition, only an upper limit was obtained.  The emitting region, offset 18\asec\ south of \molh-Peak\,1, appeared point-like in our observations with APEX.}
{The extremely high spectral line density in Orion makes the identification of \peroxyo\ much more difficult than in \roa. As a result of having to consider the possible contamination by other molecules,  we left the current detection status  undecided.}

\maketitle

%
%
\section{Introduction}

Searches for the presumed key molecule \molo\ \citep{goldsmith1978} in numerous star-forming regions have been highly unawarding \citep[e.g.][]{goldsmith2000,pagani2003}, with the definite detection of the molecule in merely two sources, viz. \roa\ \citep{larsson2007,liseau2012} and Orion\,A \citep{goldsmith2011,chen2014}. Some cases have been either resolved or remained undecided \citep[e.g.][]{goldsmith2002,yildiz2013}.

The observed scarcity of \molo\ in the  Interstellar Medium (ISM) called for the abandonment of pure gas-phase chemistry models and the invocation of grain-surface processes \citep{hollenbach2009}. Specific models addressed the conditions of the Orion Bar PDR (photodissociation region), where searches had however been unsuccessful in detecting the molecule \citep{melnick2012}. Surprisingly, perhaps, \molo\ was detected towards the hot core, albeit at an LSR (Local Standard of Rest)-velocity of 10-12\,\kms, i.e., significantly different from that of typical hot core molecules \citep[\about\,5\,\kms;][and references therein]{goddi2011}. These authors also found a small region of emission in \ammonia\ inversion lines with velocities of about 11\,kms. Overall, line widths decrease with excitation from \about\,5\,\kms\ to \about\,2\,\kms.

\citet{chen2014} were able to pinpoint the location of the 9\asec\ \molo\ source, near the position identified as \molh-Peak\,1 and somewhat offset from the hot core centre. The non-detection of the \molo\ line at 1121\,GHz  led the authors to conclude that gas temperatures do not exceed 50\,K, with best-fit model values more like 30\,K. The excitation conditions thus resemble those in \roa\ \citep{liseau2012}.

\citet{du2012} developed models for grain surface chemistry, and as an example,  they considered the particular case of \roa. According to these models, the existence of \molo\ in the gas phase is a transient phenomenon, lasting for some \powten{5} years, and which may explain the extremely few detections. These models also predict the accompanying occurrence of hydrogen peroxide (\peroxyo\ or \peroxy) and hydroperoxyl (HO$_2$), and water of course, via the following major reactions on grain surfaces \citep{tielens1982, parise2014}:

\begin{center}
\molo\ + H $\rightarrow$ H\molo \\ 
\vspace{0.2cm}

H\molo\ + H $\rightarrow$ \peroxyo \\
\vspace{0.2cm}

\peroxyo\ + H $\rightarrow$ \water\ + OH, 
\end{center}

\noindent
and these two species were then also firstly detected in \roa\ \citep{bergman2011a,parise2012}. As was the case with \molo, the observation of  ten other targets in lines of \peroxyo\ gave null results \citep{parise2014}, supporting the \molo-\peroxyo\ association. This included low- and high-mass star formation regions, where in particular the high-mass star formation regions had strong UV fields, shocks and maser emissions. It was natural, therefore, to search for the hydrogen peroxide molecule in Orion\,A, a site that was not listed in Table\,4 of \citet{parise2014}.

The organisation of this {\it Research Note} is briefly outlined as follows: in Sect.\,2, the observations and data reduction are reported, with the results provided in Sect.\,3. A brief discussion, together with our conclusions, follows in Sect.\,4. 

%
%
\section{Observations and data reduction}

The region around the position ``Orion \molh-Peak\,1" \citep{chen2014} was observed with the Atacama Pathfinder Experiment (APEX; {\small \texttt{http://www.apex-telescope.org/}}) in 2014 during the time August to December (Table\,\ref{obs_log}).  APEX is a 12\,m single dish telescope at 5100\,m altitude in northern Chile. We used two receivers from the SHeFI\footnote{Swedish Heterodyne Facility Instrument} suite, i.e., APEX-1 for $(3_{03}-2_{11})$\,219\,GHz and $(6_{15}-5_{05})$\,252\,GHz  and APEX-2 for $(4_{04}-3_{12})$\,269\,GHz and $(5_{05}-4_{13})$, $(5_{14}-6_{06})$\,319\,GHz, respectively\footnote{Energy level diagrams are found in \citet{bergman2011a}.}. At these frequencies, the HPBW of APEX is 20\asec\ to 28\asec. The rms value of the telescope pointing accuracy is 2\asec.

As seen in Table\,\ref{obs_log}, maps were obtained on-the-fly in the 219\,GHz and 252\,GHz lines, with a sampling rate of 9\asec/pxl, oversampling the $3^{\prime} \times 3^{\prime}$ region in these lines. The central J2000-coordinates are 
R.A. = \radot{05}{35}{13}{70}, Dec. = \decdot{$-05$}{22}{09}{0}. Towards the offset position (0\asec, $-18$\asec), single position spectra were obtained at 269\,GHz and 319\,GHz.

For the instantaneous bandwidth of 2.5\,GHz, we used as backend the Fast Fourier Transform Spectrometer (FFTS) with 32768 spectral channels. We selected a spectral resolution of 76.3\,kHz per channel, corresponding to a velocity resolution of \about\,0.1\,\kms. The data were reduced with the software packages GILDAS/CLASS (\texttt{\small {http://www.ira.inaf.it/$\sim$brand/gag.html}}) and \texttt{xs} \\ (\texttt{\small{ftp://yggdrasil.oso.chalmers.se/pub/xs/}}).

%
%
\section{Results}

\begin{figure*}
  \resizebox{\hsize}{!}{
    \rotatebox{0}{\includegraphics{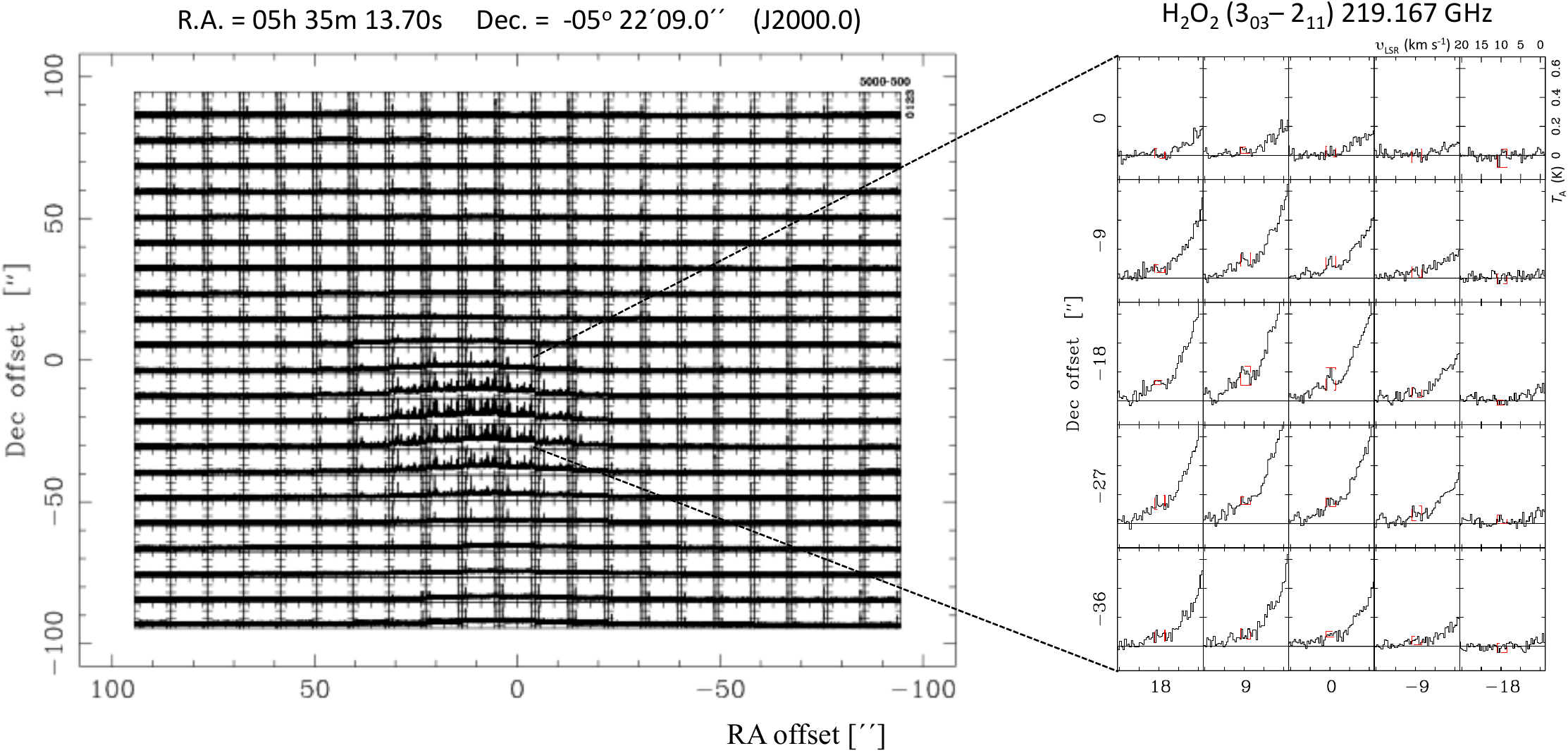}}  
                        }
  \caption{{\bf Left:} the $3^{\prime} \times 3^{\prime}$ mapped area, sampled at 9\asec\  with the origin at the  Orion \molh-Peak\,1 position, i.e., \atwozero\ = \radot{05}{35}{13}{70}, \dtwozero\, = \decdot{$-05$}{22}{09}{0}.  A core of intense emission is clearly seen just below the centre. {\bf Right:} Centred on (0\asec, $-18$\asec), this partial map demonstrates that the 219.17\,GHz feature is a point source to the 28\asec\ beam. This weak spectral feature is identified inside the red markers. It is sitting on top of the red wing of a much stronger line  \citep[HC$_3$N\,($\nu_7=3)$,][]{sutton1985}.
  }
    \label{map}
\end{figure*}

An overview of the mapped region is shown in the left panel of Fig.\,\ref{map}, revealing that the core region near the centre is very compact. A blow-up, 36\asec\ in size, is shown in the right panel, where a weak emission feature is shown on the wing of a stronger line. That feature corresponds to the ($3_{03}-2_{11}$) line of \peroxyo\ at the LSR-velocity of 10.0\,\kms, i.e., consistent with that of the \molo\ lines \citep{chen2014}. It can also be seen that this feature is not merely due to noise, but is repeatedly seen in different positions, albeit at lower intensity. The fact that HOOH is not detected outside this limited region implies that the emission in the 219\,GHz line is point-like to the 28\asec\ telescope beam. 

From the data in Table\,\ref{results}, it appears that only two out of five lines were clearly detected ($\ge 4 \sigma$), and two were possibly detected at low S/N ($2.8-3.5 \sigma$). The quoted line widths (FWHMs) are  only lower limits because of
the difficulty in accurately determining the local continuum on sloping backgrounds. These \peroxyo\ widths are smaller than those for \molo\ reported by \citet{chen2014}. The 252\,GHz line was not detected. However, the noise level of that spectrum is very much higher than for the other observations (Table\,\ref{results}).
\begin{table}
    \caption{Log of observations}
    \label{obs_log}
    \begin{tabular}{clcrl}      
\hline\hline    
\noalign{\smallskip}             \noalign{\smallskip}   
  \peroxyo-transition   & Frequency     & Date                  & $t_{\rm int}$   & Sp.    \\
$(J_{K_a K_c})^{\prime} - (J_{K_a K_c})^{\prime \prime}$        &  (GHz)        &  {\small yy--mm--dd}    & (min) &       \\ 
\noalign{\smallskip}    
\hline                          
\noalign{\smallskip}    
$3_{03}-2_{11}$ & 219.16686     & 14-08-14              & 60.0          & map     \\
                                &                       & 14-08-18              & 60.0            & map   \\
                                &                       & 14-08-19              & 60.0            & map   \\                              
                                &                       & 14-10-06              & 88.0            & sngl  \\      
$6_{15}-5_{05}$ & 251.91468     & 14-08-15              & 60.0          & map     \\
$4_{04}-3_{12}$ & 268.96117     & 14-10-06              & 73.1          & sngl    \\
                                &                       & 14-10-08              & 21.9            & sngl  \\
                                &                       & 14-12-04              & 167.9           & sngl  \\
$5_{05}-4_{13}$ & 318.22325     & 14-10-08              & 58.6          & sngl    \\
                                &                       & 14-10-10              & 40.3            & sngl  \\      
$5_{14}-6_{06}$ & 318.71210     & 14-10-08              & 58.6          & sngl    \\
                                &                       & 14-10-10              & 40.3            & sngl  \\              
\noalign{\smallskip}    
\hline
    \end{tabular}
    \begin{list}{}{}
    \item[] Maps have their origin at \atwozero\,= \radot{05}{35}{13}{70}, \dtwozero\,= \decdot{$-05$}{22}{09}{0} \citep{chen2014}, and single position spectra refer to the offset (0\asec, $-18$\asec).
    \end{list}
\end{table}

\begin{table}
    \caption{Measurements of \peroxyo\ features}
    \label{results}
    \begin{tabular}{ccccrcc}      
\hline\hline    
\noalign{\smallskip}             \noalign{\smallskip}   
  \peroxyo                              & $E_{\rm up}/k$        & \vlsr &  FWHM   &  \trms        & \tadv         & S/N\\
Line                                    &               (K)             &   (\kms)        &  (\kms)       &  (mK)         &(K \kms)       &       \\ 
\noalign{\smallskip}    
\hline                          
\noalign{\smallskip}    
$3_{03}-2_{11}$         & 31                            &  10.0 & 1.4           & 26.1    & 0.172                 & 4.6   \\
$6_{15}-5_{05}$         & 66                            & 11.6  & 0.8           &257.0  & 0.242           & 1.2   \\
$4_{04}-3_{12}$         & 41                            &  10.3 & 1.0           & 41.9    & 0.118         & 2.8\\
$5_{05}-4_{13}$         & 53                            &  11.8 & 0.9           & 28.8    & 0.102         & 3.5   \\
$5_{14}-6_{06}$         & 67                            &  12.3 & 1.5           & 31.5    & 0.174         & 4.0   \\      
\noalign{\smallskip}    
\hline
    \end{tabular}
\end{table}

%
%
\section{Discussion and conclusions}

\begin{figure*}
  \resizebox{\hsize}{!}{
    \rotatebox{0}{\includegraphics{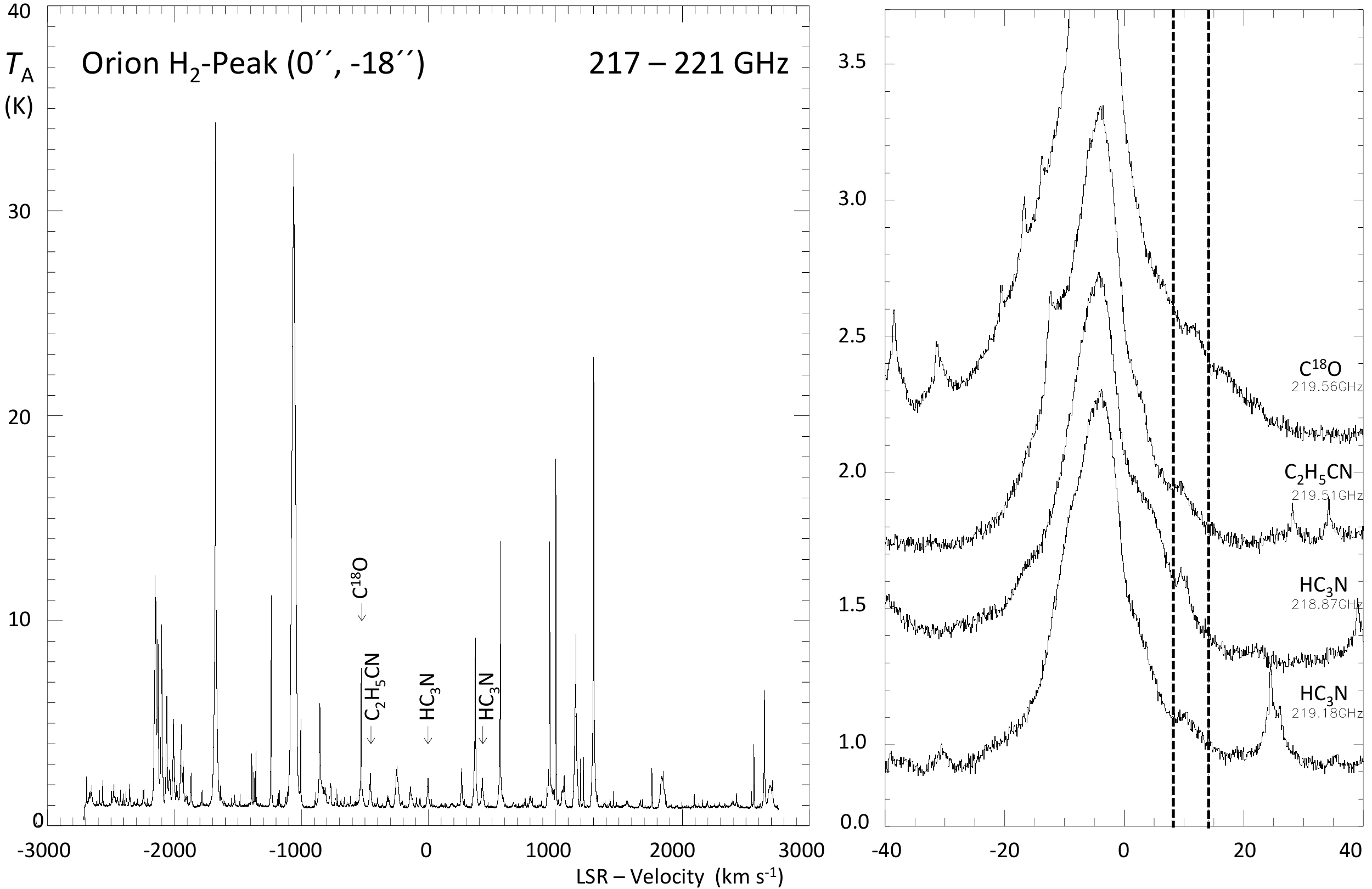}}  
                        }
  \caption{{\bf Left:}  The 4 GHz wide spectrum, centred on 219\,GHz, towards the offset position (0\asec, $-18$\asec) relative to Orion \molh-Peak\,1. Line identifications for the entire spectral region can be found in the paper by \citet{sutton1985}. Blow-ups of the labelled lines are found in the {\bf right}-hand panel, where the LSR-velocity range of the putative \peroxyo\  line is indicated with the dashed vertical lines.
   }
  \label{lines}
\end{figure*}

The LSR-velocities of the \peroxyo\ features are clearly outside the hot core window, but seem consistent with those obtained for \molo. This could also indicate that  in Orion \peroxyo\ can be tied to \molo. A major shortcoming, though, is the extremely high line density towards the hot core region, which makes proper line identification  difficult. In fact,  several molecules in the 219\,GHz spectrum display similar hump features on their red wings (Fig.\,\ref{lines}). This is not evidenced by the other transitions, but in view of the relatively lower S/N makes the \peroxyo\ identification apparently  non-unique.

\begin{acknowledgement}  The contributions by P.\,Bergman, including the interesting discussions, are highly appreciated. We also thank the Swedish APEX team and the APEX staff on site for their help with the observations. As part of our Odin and {\it Herschel} work, this research has been supported by the Swedish National Space Board (SNSB).
\end{acknowledgement}

\bibliographystyle{aa}
\bibliography{refs}

\begin{thebibliography}{18}
\expandafter\ifx\csname natexlab\endcsname\relax\def\natexlab#1{#1}\fi

\bibitem[{{Bergman} {et~al.}(2011){Bergman}, {Parise}, {Liseau}, {Larsson},
  {Olofsson}, {Menten}, \& {G{\"u}sten}}]{bergman2011a}
{Bergman}, P., {Parise}, B., {Liseau}, R., {et~al.} 2011, \aap, 531, L8

\bibitem[{{Chen} {et~al.}(2014){Chen}, {Goldsmith}, {Viti}, {Snell}, {Lis},
  {Benz}, {Bergin}, {Black}, {Caselli}, {Encrenaz}, {Falgarone}, {Goicoechea},
  {Hjalmarson}, {Hollenbach}, {Kaufman}, {Melnick}, {Neufeld}, {Pagani}, {van
  der Tak}, {van Dishoeck}, \& {Y{\i}ld{\i}z}}]{chen2014}
{Chen}, J.-H., {Goldsmith}, P.~F., {Viti}, S., {et~al.} 2014, \apj, 793, 111

\bibitem[{{Du} {et~al.}(2012){Du}, {Parise}, \& {Bergman}}]{du2012}
{Du}, F., {Parise}, B., \& {Bergman}, P. 2012, \aap, 538, A91

\bibitem[{{Goddi} {et~al.}(2011){Goddi}, {Greenhill}, {Humphreys}, {Chandler},
  \& {Matthews}}]{goddi2011}
{Goddi}, C., {Greenhill}, L.~J., {Humphreys}, E.~M.~L., {Chandler}, C.~J., \&
  {Matthews}, L.~D. 2011, \apjl, 739, L13

\bibitem[{{Goldsmith} \& {Langer}(1978)}]{goldsmith1978}
{Goldsmith}, P.~F. \& {Langer}, W.~D. 1978, \apj, 222, 881

\bibitem[{{Goldsmith} {et~al.}(2002){Goldsmith}, {Li}, {Bergin}, {Melnick},
  {Tolls}, {Howe}, {Snell}, \& {Neufeld}}]{goldsmith2002}
{Goldsmith}, P.~F., {Li}, D., {Bergin}, E.~A., {et~al.} 2002, \apj, 576, 814

\bibitem[{{Goldsmith} {et~al.}(2011){Goldsmith}, {Liseau}, {Bell}, {Black},
  {Chen}, {Hollenbach}, {Kaufman}, {Li}, {Lis}, {Melnick}, {Neufeld}, {Pagani},
  {Snell}, {Benz}, {Bergin}, {Bruderer}, {Caselli}, {Caux}, {Encrenaz},
  {Falgarone}, {Gerin}, {Goicoechea}, {Hjalmarson}, {Larsson}, {Le Bourlot},
  {Le Petit}, {De Luca}, {Nagy}, {Roueff}, {Sandqvist}, {van der Tak}, {van
  Dishoeck}, {Vastel}, {Viti}, \& {Y{\i}ld{\i}z}}]{goldsmith2011}
{Goldsmith}, P.~F., {Liseau}, R., {Bell}, T.~A., {et~al.} 2011, \apj, 737, 96

\bibitem[{{Goldsmith} {et~al.}(2000){Goldsmith}, {Melnick}, {Bergin}, {Howe},
  {Snell}, {Neufeld}, {Harwit}, {Ashby}, {Patten}, {Kleiner}, {Plume},
  {Stauffer}, {Tolls}, {Wang}, {Zhang}, {Erickson}, {Koch}, {Schieder},
  {Winnewisser}, \& {Chin}}]{goldsmith2000}
{Goldsmith}, P.~F., {Melnick}, G.~J., {Bergin}, E.~A., {et~al.} 2000, \apjl,
  539, L123

\bibitem[{{Hollenbach} {et~al.}(2009){Hollenbach}, {Kaufman}, {Bergin}, \&
  {Melnick}}]{hollenbach2009}
{Hollenbach}, D., {Kaufman}, M.~J., {Bergin}, E.~A., \& {Melnick}, G.~J. 2009,
  \apj, 690, 1497

\bibitem[{{Larsson} {et~al.}(2007){Larsson}, {Liseau}, {Pagani}, {Bergman},
  {Bernath}, {Biver}, {Black}, {Booth}, {Buat}, {Crovisier}, {Curry},
  {Dahlgren}, {Encrenaz}, {Falgarone}, {Feldman}, {Fich}, {Flor{\'e}n},
  {Fredrixon}, {Frisk}, {Gahm}, {Gerin}, {Hagstr{\"o}m}, {Harju}, {Hasegawa},
  {Hjalmarson}, {Johansson}, {Justtanont}, {Klotz}, {Kyr{\"o}l{\"a}}, {Kwok},
  {Lecacheux}, {Liljestr{\"o}m}, {Llewellyn}, {Lundin}, {M{\'e}gie},
  {Mitchell}, {Murtagh}, {Nordh}, {Nyman}, {Olberg}, {Olofsson}, {Olofsson},
  {Olofsson}, {Persson}, {Plume}, {Rickman}, {Ristorcelli}, {Rydbeck},
  {Sandqvist}, {Sch{\'e}ele}, {Serra}, {Torchinsky}, {Tothill}, {Volk},
  {Wiklind}, {Wilson}, {Winnberg}, \& {Witt}}]{larsson2007}
{Larsson}, B., {Liseau}, R., {Pagani}, L., {et~al.} 2007, \aap, 466, 999

\bibitem[{{Liseau} {et~al.}(2012){Liseau}, {Goldsmith}, {Larsson}, {Pagani},
  {Bergman}, {Le Bourlot}, {Bell}, {Benz}, {Bergin}, {Bjerkeli}, {Black},
  {Bruderer}, {Caselli}, {Caux}, {Chen}, {de Luca}, {Encrenaz}, {Falgarone},
  {Gerin}, {Goicoechea}, {Hjalmarson}, {Hollenbach}, {Justtanont}, {Kaufman},
  {Le Petit}, {Li}, {Lis}, {Melnick}, {Nagy}, {Olofsson}, {Olofsson}, {Roueff},
  {Sandqvist}, {Snell}, {van der Tak}, {van Dishoeck}, {Vastel}, {Viti}, \&
  {Y{\i}ld{\i}z}}]{liseau2012}
{Liseau}, R., {Goldsmith}, P.~F., {Larsson}, B., {et~al.} 2012, \aap, 541, A73

\bibitem[{{Melnick} {et~al.}(2012){Melnick}, {Tolls}, {Goldsmith}, {Kaufman},
  {Hollenbach}, {Black}, {Encrenaz}, {Falgarone}, {Gerin}, {Hjalmarson}, {Li},
  {Lis}, {Liseau}, {Neufeld}, {Pagani}, {Snell}, {van der Tak}, \& {van
  Dishoeck}}]{melnick2012}
{Melnick}, G.~J., {Tolls}, V., {Goldsmith}, P.~F., {et~al.} 2012, \apj, 752, 26

\bibitem[{{Pagani} {et~al.}(2003){Pagani}, {Olofsson}, {Bergman}, {Bernath},
  {Black}, {Booth}, {Buat}, {Crovisier}, {Curry}, {Encrenaz}, {Falgarone},
  {Feldman}, {Fich}, {Floren}, {Frisk}, {Gerin}, {Gregersen}, {Harju},
  {Hasegawa}, {Hjalmarson}, {Johansson}, {Kwok}, {Larsson}, {Lecacheux},
  {Liljestr{\"o}m}, {Lindqvist}, {Liseau}, {Mattila}, {Mitchell}, {Nordh},
  {Olberg}, {Olofsson}, {Ristorcelli}, {Sandqvist}, {von Scheele}, {Serra},
  {Tothill}, {Volk}, {Wiklind}, \& {Wilson}}]{pagani2003}
{Pagani}, L., {Olofsson}, A.~O.~H., {Bergman}, P., {et~al.} 2003, \aap, 402,
  L77

\bibitem[{{Parise} {et~al.}(2012){Parise}, {Bergman}, \& {Du}}]{parise2012}
{Parise}, B., {Bergman}, P., \& {Du}, F. 2012, \aap, 541, L11

\bibitem[{{Parise} {et~al.}(2014){Parise}, {Bergman}, \& {Menten}}]{parise2014}
{Parise}, B., {Bergman}, P., \& {Menten}, K. 2014, ArXiv e-prints
  [\eprint[arXiv]{1407.0550}]

\bibitem[{{Sutton} {et~al.}(1985){Sutton}, {Blake}, {Masson}, \&
  {Phillips}}]{sutton1985}
{Sutton}, E.~C., {Blake}, G.~A., {Masson}, C.~R., \& {Phillips}, T.~G. 1985,
  \apjs, 58, 341

\bibitem[{{Tielens} \& {Hagen}(1982)}]{tielens1982}
{Tielens}, A.~G.~G.~M. \& {Hagen}, W. 1982, \aap, 114, 245

\bibitem[{{Y{\i}ld{\i}z} {et~al.}(2013){Y{\i}ld{\i}z}, {Acharyya}, {Goldsmith},
  {van Dishoeck}, {Melnick}, {Snell}, {Liseau}, {Chen}, {Pagani}, {Bergin},
  {Caselli}, {Herbst}, {Kristensen}, {Visser}, {Lis}, \& {Gerin}}]{yildiz2013}
{Y{\i}ld{\i}z}, U.~A., {Acharyya}, K., {Goldsmith}, P.~F., {et~al.} 2013, \aap,
  558, A58

\end{thebibliography}


\end{document}